\documentclass{article}
\usepackage{spconf,amsmath,graphicx}
\usepackage{subfig}
\usepackage{amsfonts}
\usepackage{multirow}
\PassOptionsToPackage{hyphens}{url}\usepackage{hyperref}

\title{Real-time Wireless ECG-derived Respiration Rate Estimation Using an Autoencoder with a DCT Layer}
\name{Hongyi Pan* \quad Xin Zhu* \quad Zhilu Ye*\quad Pai-Yen Chen\quad Ahmet Enis Cetin\thanks{This work was supported by National Science Foundation (NSF) under grant 1934915. Xin Zhu was supported  by NSF IDEAL 2217023.}}
\address{Department of Electrical and Computer Engineering, University of Illinois Chicago, Chicago, IL}
\begin{document}

%
\maketitle
\def\thefootnote{*}\footnotetext{These authors contributed equally to this work.
Our code is available at \url{https://github.com/phy710/ICASSP2023-AE-DCT}.}\def\thefootnote{\arabic{footnote}}
\begin{abstract}

In this paper, we present a wireless ECG-derived Respiration Rate (RR) estimation using an autoencoder with a DCT Layer. The wireless wearable system records the ECG data of the subject and the respiration rate is determined from the variations in the baseline level of the ECG data. A straightforward Fourier analysis of the ECG data obtained using the wireless wearable system may lead to incorrect results due to uneven breathing. To improve the estimation precision, we propose a neural network that uses a novel Discrete Cosine Transform (DCT) layer to denoise and decorrelates the data. The DCT layer has trainable weights and soft-thresholds in the transform domain. In our dataset, we improve the Mean Squared Error (MSE) and Mean Absolute Error (MAE) of the Fourier analysis-based approach using our novel neural network with the DCT layer.
\end{abstract}
\section{Introduction}
\label{sec:intro}
It is well-known that breathing is an essential physiological task in living organisms. Respiration rate is one of the four vital signs routinely monitored by health care providers and medical staffs~\cite{al2011respiration}. There are various respiration monitoring methods based on acoustic~\cite{moussavi2000computerised,muthusamy2020computerized}, airflow~\cite{muthusamy2019overview}, chest and abdominal movement~\cite{stromberg1998thoracoabdominal,nepal2002apnea}, oximetry probe (SpO$_2$)~\cite{leonard2003standard}, infrared vibration sensors~\cite{erden2015respiratory,erden2017period,cetin2021review, erden2015contact, kapu2017resting, erden2016sensors}, while the Electrocardiogram (ECG) based respiration monitoring methods are one of the most effective methods~\cite{mazzanti2003validation,dingab2004derivation,tarassenko2002multi}.
In the ECG-based respiration rate monitoring approaches, ECG electrodes are attached to the subject to record ECG measurements~\cite{moody1986clinical}. Traditional ECG sensing systems require wire connections between the test equipment and the human body, which greatly restricts the patient's movements and they are uncomfortable. Patients typically rely on medical professionals to monitor their heart condition, which demands repeated trips to the hospital or long-term hospitalization. Therefore, there is a great demand to establish a wireless ECG measuring system, which can record both the ECG and the Respiration Rate (RR) of the subject.

It is important to  continually monitor the RR  \cite{nicolo2020importance, marjanovic2020easy} because the RR is one of the best indicators of clinical deterioration of health. Unfortunately, the measurements are often manually performed by nurses and they are not accurate. This paper provides a practical solution to RR monitoring problem by estimating the RR using a wearable wireless ECG system.

In this paper, we present an autoencoder-type neural network with a novel transform domain layer to accurately determine the respiration rate. We use the Fourier transform of the ECG signals obtained from the wireless ECG monitoring system 
as the input to the network. The network makes a decision using the DFT magnitudes as its input. The transform domain layer of the network is based on the discrete cosine transform (DCT) and it uses a trainable soft-thresholding function to denoise the data. The network learns the thresholds using the backpropagation algorithm.

\begin{figure}[htbp]
    \centering
    \includegraphics[width=1\linewidth]{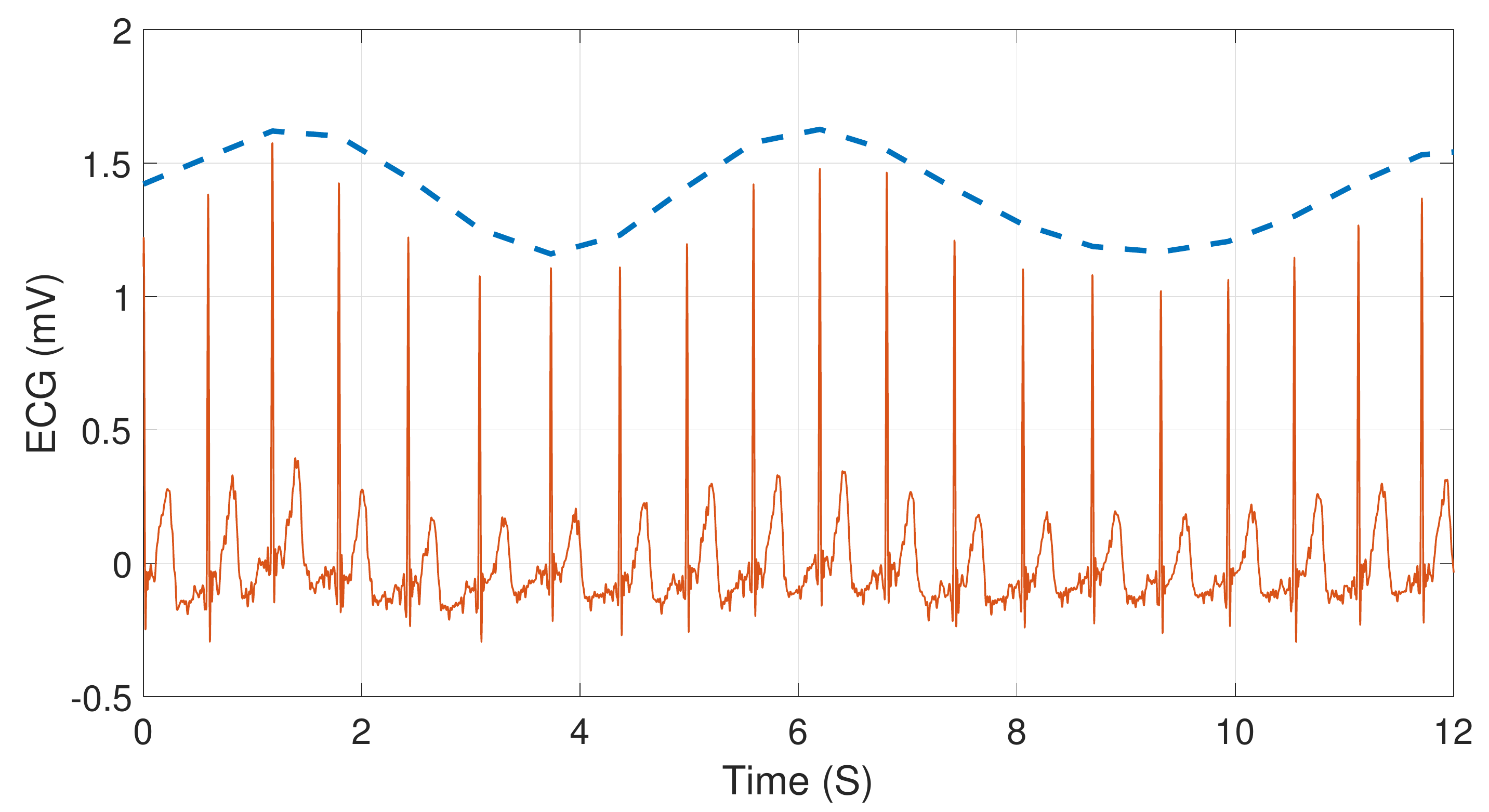}
    \caption{An ECG recording. The baseline level of the ECG waveform varies due to the respiration activity of the subject as indicated by the dashed line (manually drawn by us).}
    \label{fig: ECG}
\end{figure}

Our contributions are summarized as the following:
 The baseline level of the ECG data moves due to chest motion as shown in Fig.~\ref{fig: ECG} but this data is very noisy. Traditional Fourier analysis may lead to incorrect results due to the multiple and closely located peaks in the Fourier domain. To solve this problem we introduce a neural network with a DCT domain layer, which has trainable weights and soft thresholds. Since the DCT is real it can be easily used as a part of a neural network.  The network with a DCT layer has better accuracy than a conventional neural network.
 
\section{Methodology}
In this section, we first briefly introduce our wireless ECG monitoring system for data collecting. Then we will describe how we design an autoencoder and our novel DCT layer to assist the traditional Fourier analysis to improve the accuracy of the respiration rate estimation process.

\subsection{Wireless ECG monitoring system}
\begin{figure}[htbp]
     \centering
     \includegraphics[width=1\linewidth]{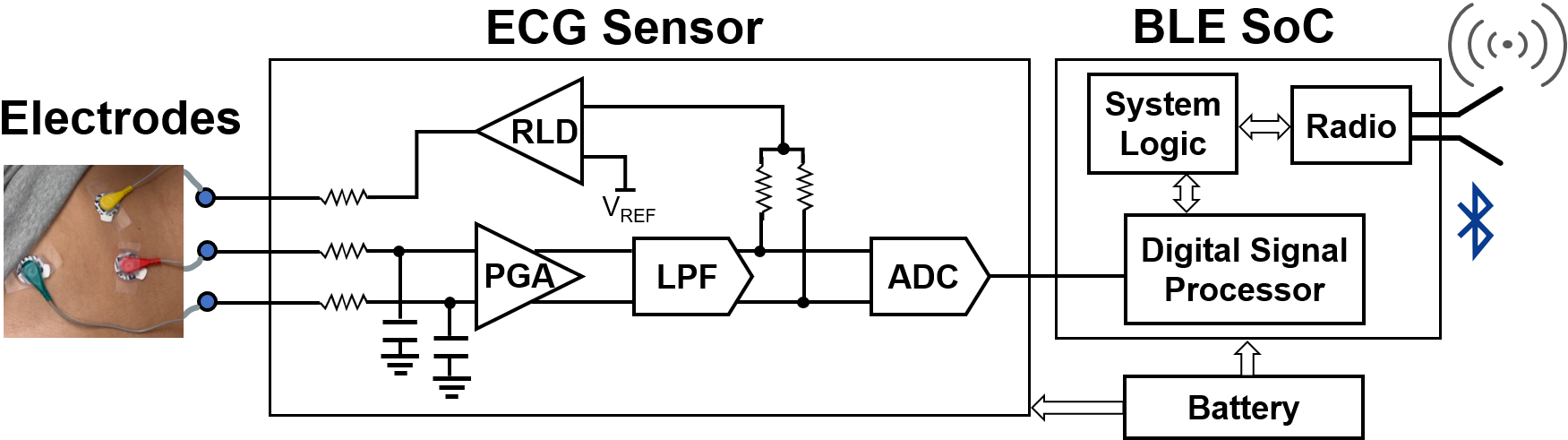}
     \caption{The block diagram of the wireless ECG monitoring system. Three ECG electrodes are attached to the body. }
     \label{fig: system}
 \end{figure}

 We use a wireless ECG monitoring system to collect the ECG data from the chest area of the subject. 
 Three ECG electrodes are  attached to the body as shown in Fig.~\ref{fig: system}. A recorded ECG waveform is shown in Fig.~\ref{fig: ECG}. The data is transmitted to a base station using a wireless system. The baseline level of the ECG waveform varies due to the respiration activity of the subject as indicated by the dashed line (manually drawn by us). 
 
\subsection{Fourier Analysis of the ECG data for RR Estimation}\label{sec:Fourier Analysis Method}
\begin{figure}[htbp]
\centering
\subfloat[DFT detects the true RR.\\ DFT estimated rate is 19 bpm.]{\includegraphics[width=0.5\linewidth]{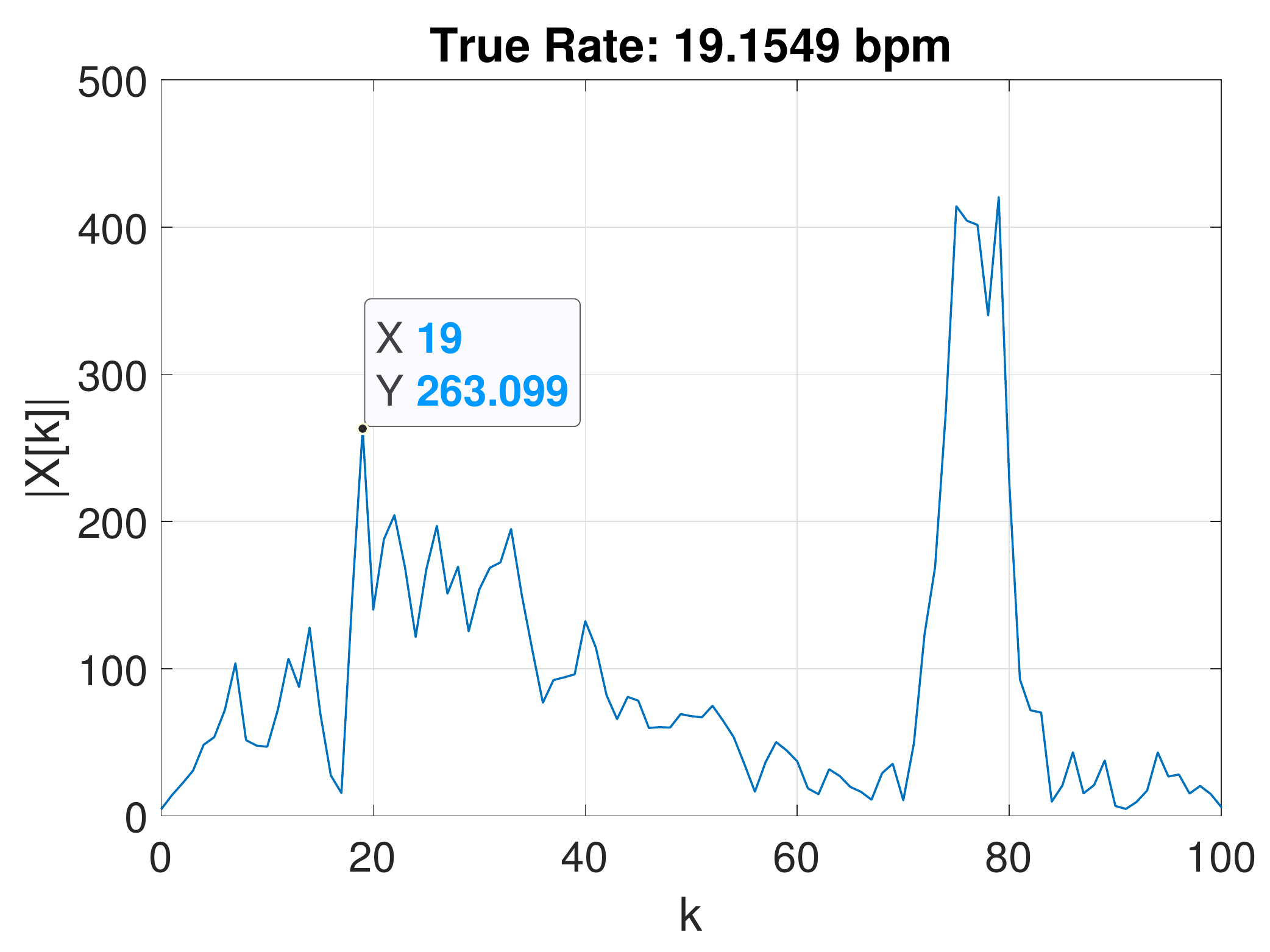}}
\subfloat[DFT detects the true RR.\\ DFT estimated rate is 23 bpm.]{\includegraphics[width=0.5\linewidth]{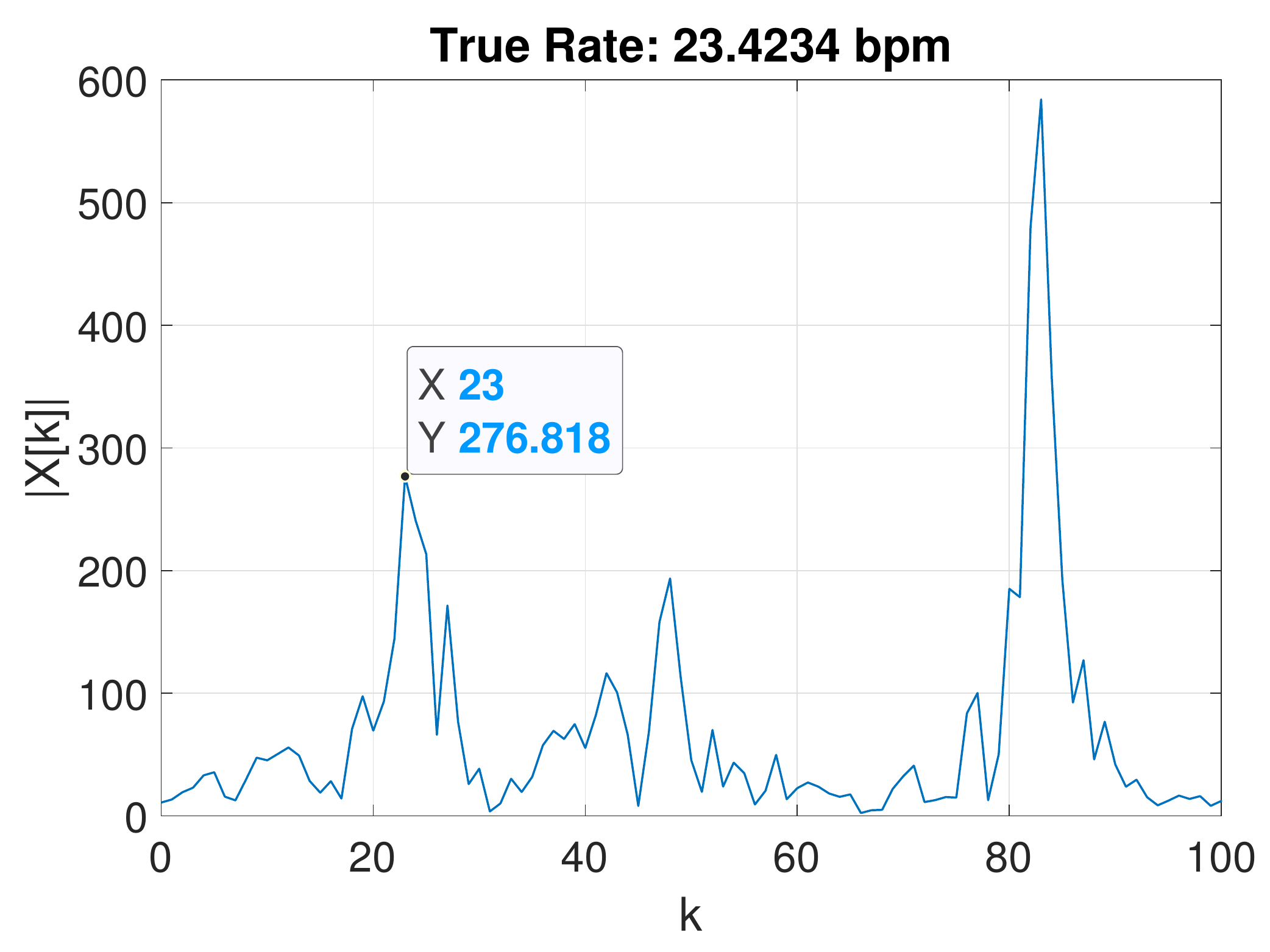}}\\
\subfloat[DFT fails to determine \\ the true RR=20 bpm.]{\includegraphics[width=0.5\linewidth]{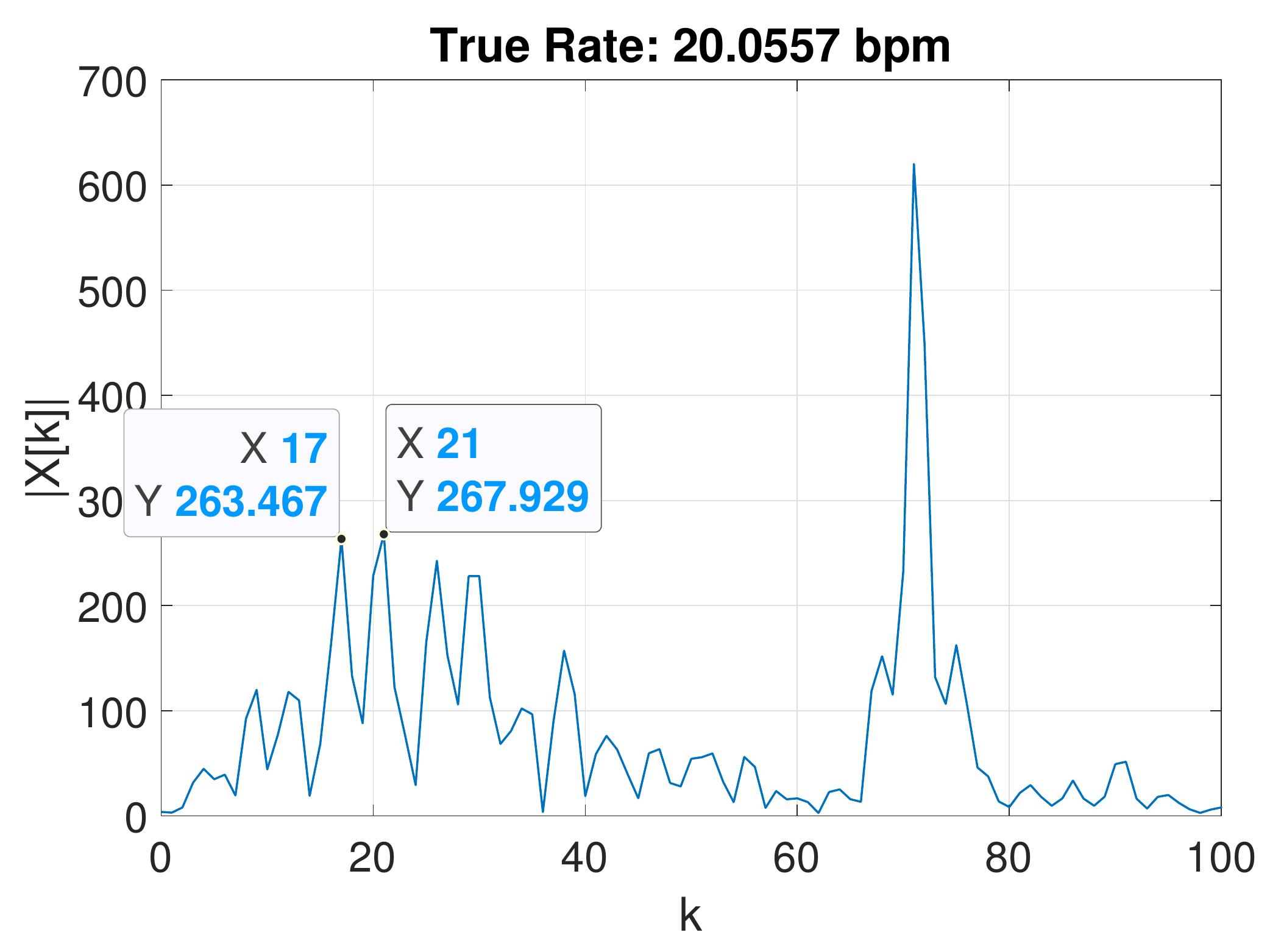}}
\subfloat[DFT fails to determine\\ the true RR=18 bpm.]{\includegraphics[width=0.5\linewidth]{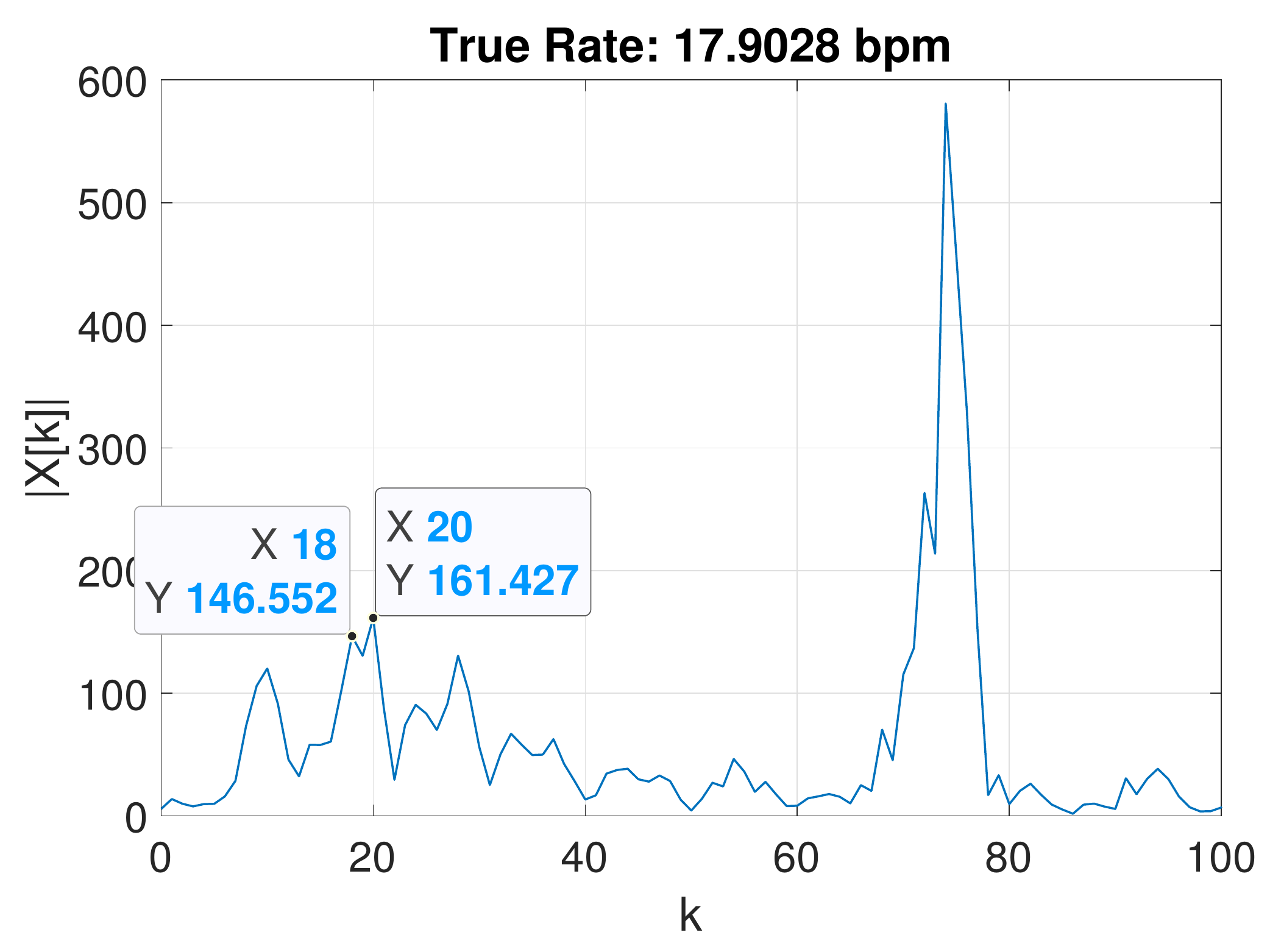}}
\caption{Discrete Fourier Transform (DFT) of four different ECG waveforms. In each figure, we find the largest peak from the first 27 DFT entries. The highest peak below k=27 breaths per minute (bpm) may represent the true Respiration Rate (RR). However, we have incorrect results in (c) and (d).}
\label{fig: results}
\end{figure}

We divide the ECG data into short-time windows, then apply a Hamming window and calculate the Discrete Fourier Transform (DFT) of each window for reparation rate estimation. The duration of the short-time window is one minute.
In Fig.~\ref{fig: results} we have four DFT magnitude plots. We use the DFT data up to the DFT index $k=27$ in Fig.~\ref{fig: results} (the index starts from 0) because the respiration rate is normally less than 26 breaths per minute. As can be seen from 
Fig.~\ref{fig: results} DFT plots have multiple peaks and the peaks are too close to each other. In the top row of  Fig.~\ref{fig: results}, we observe clear peaks corresponding to the correct respiration rates. However, the DFT fails to produce clear peaks below $k=27$ in the bottom row of Fig.~\ref{fig: results}. This is because people do not always breathe in a perfect periodical manner. Therefore, we do not have a single peak in the Fourier domain.
\subsection{Autoencoder Network for RR Estimation}
As pointed out above, we cannot observe a single clear peak using the Fourier analysis in many cases. Therefore, finding the index of the maximum value in the Fourier spectrum directly is insufficient for predicting the respiration rate accurately. In this section, we design an autoencoder network~\cite{rumelhart1985learning} to estimate the respiration rate. The input to the autoencoder is the DFT magnitude vector and it is trained using the Gaussian function. We would like to have a single peak in the respiration rate range of $[10, 26]$ breaths per minute (bpm). During training, we placed a single Gaussian whose mean is the same as the true respiration rate.

\begin{figure}[htbp]
\centering
\subfloat[\label{fig: autoencoder}Autoencoder.]{\includegraphics[width=0.27\linewidth]{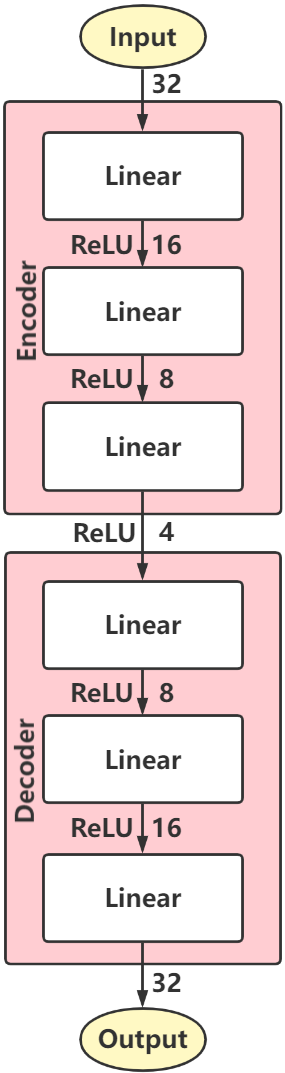}}\hspace{1pt}
\subfloat[\label{fig: dct} DCT Layer.]{\includegraphics[width=0.38\linewidth]{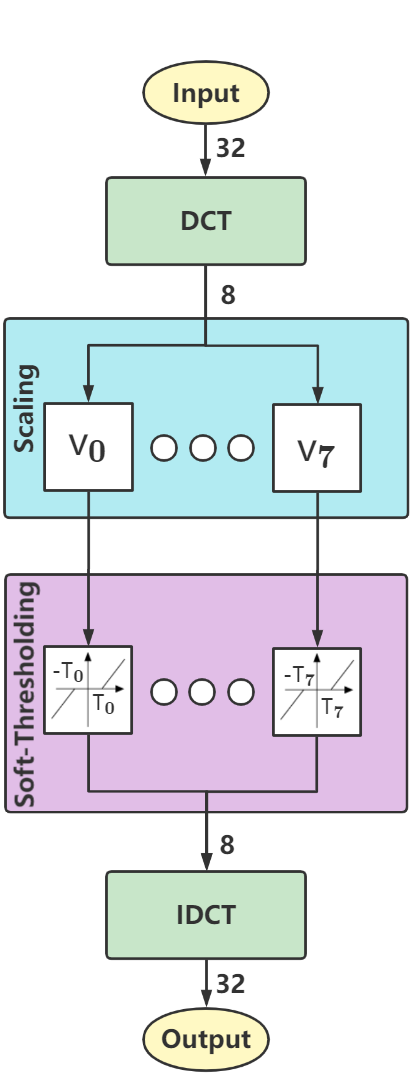}}
\caption{The block diagrams of the autoencoder and the DCT layer. The numbers after each block denote the length of the tensors (vectors). The output of the autoencoder is fed to the input of the DCT layer in Sec.~\ref{sec: Autoencoder Network with a DCT Layer}.}
\end{figure}

The autoencoder network is designed as shown in Fig.~\ref{fig: autoencoder}. It generates a 32-length pseudo-spectrum with only one peak from the first 32 points of the DFT magnitudes. Here, we use 32 instead of $k=27$ points to cover a wider range and we do not want to truncate the Gaussians corresponding to the high respiration rates.
There are three linear layers in the encoder module and three linear layers in the decoder module. The linear layer transforming the input of $\mathbf{x}\in\mathbb{R}^m$ to the output of $\mathbf{y}\in\mathbb{R}^n$ is computed as
\begin{equation}
    \mathbf{y}=\mathbf{Wx}+\mathbf{b},
\end{equation}
where, $\mathbf{W}\in\mathbb{R}^{n\times m}$ is the weight matrix and $\mathbf{b}\in\mathbb{R}^n$ is the bias vector. After these linear layers except the final one, the rectified linear unit (ReLU) function, $(x)_+=\max(x, 0)$,
~\cite{fukushima1975cognitron} is applied as the activation function. In summary, tensors in the autoencoder go as $\mathbb{R}^{32}\rightarrow\mathbb{R}^{16}\rightarrow\mathbb{R}^8\rightarrow\mathbb{R}^4\rightarrow\mathbb{R}^8\rightarrow\mathbb{R}^{16}\rightarrow\mathbb{R}^{32}$. 
The sigmoid function, $s(x)=\frac{1}{1+\exp(-x)}$, is applied after the autoencoder output to constrain the output into the range of $[0, 1]$.

To train the autoencoder network, we truncate the Fourier magnitudes into a 32-length vector as the training input for each sample because the respiration rate should be less than 32. Then, as the training desired output for each sample, we calculate a vector whose entries follow the Gaussian distribution function, $g(x)=\exp(-\frac{(x-\mu)^2}{\sigma^2})$, for $x=0, 1, 2, ..., 31$. We use the true respiration rate as the mean $\mu$ and make $\sigma =2$. We choose $\sigma =2$ because the standard deviations of true peaks are also approximately equal to 2. 
We normalize the input vector values such that the peak has a value of 1. We also normalize the Gaussian distribution function, $g(x)=\exp(-\frac{(x-\mu)^2}{\sigma^2})$, for $x=0, 1, 2, ..., 31$, during the training as the desired output for each sample. We use the true respiration rate as $\mu$ and make $\sigma =2$. 
\begin{figure}[t]
\centering
\subfloat[\label{fig: output a}DFT: 13 bpm,\\
Autoencoder: 13 bpm,\\ 
Autoencoder + DCT: 13 bpm.]{\includegraphics[width=0.5\linewidth]{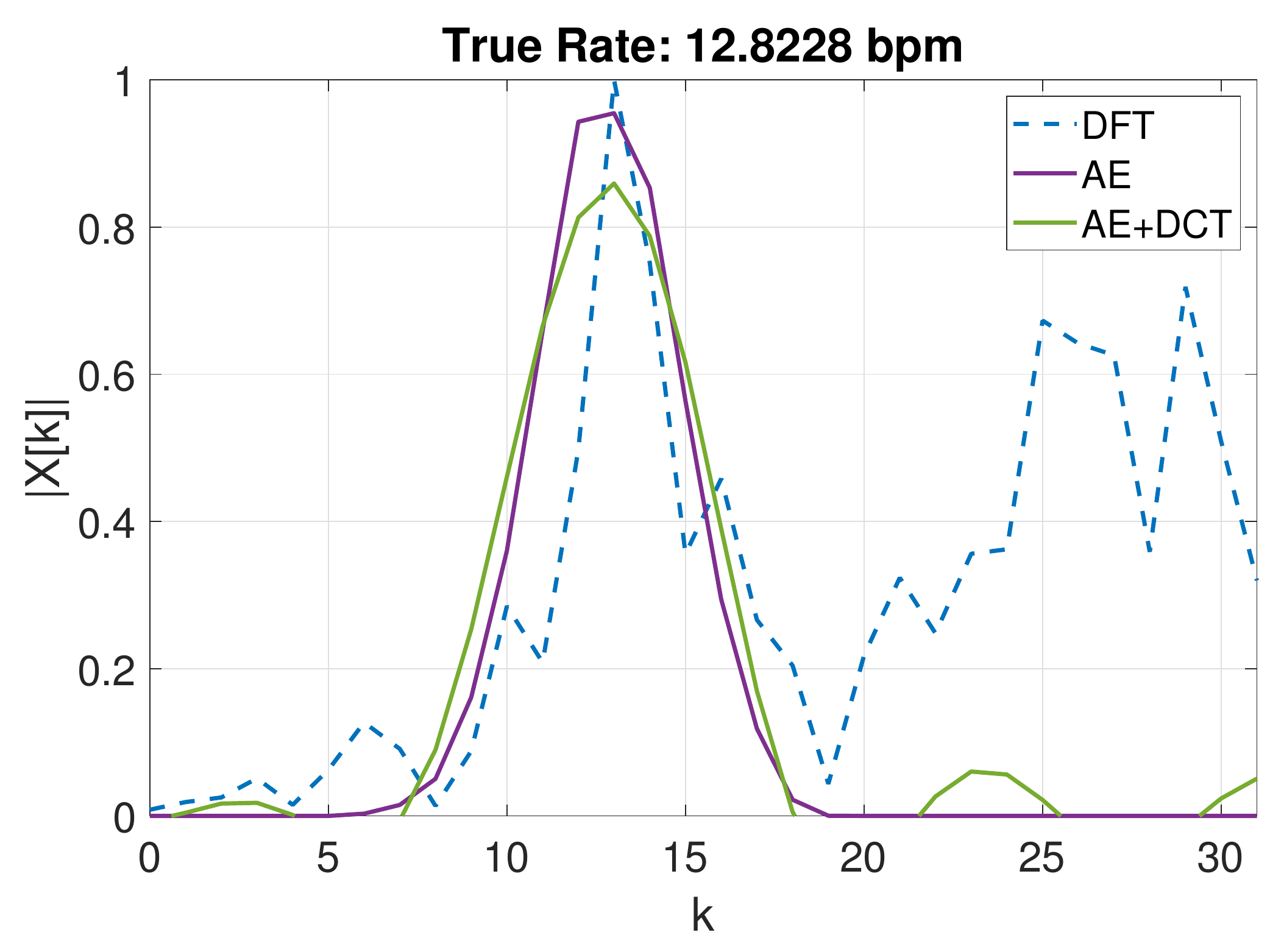}}
\subfloat[\label{fig: output b}DFT: 20 bpm or 22 bpm, \\Autoencoder: 19 bpm,\\ Autoencoder + DCT: 19 bpm.]{\includegraphics[width=0.5\linewidth]{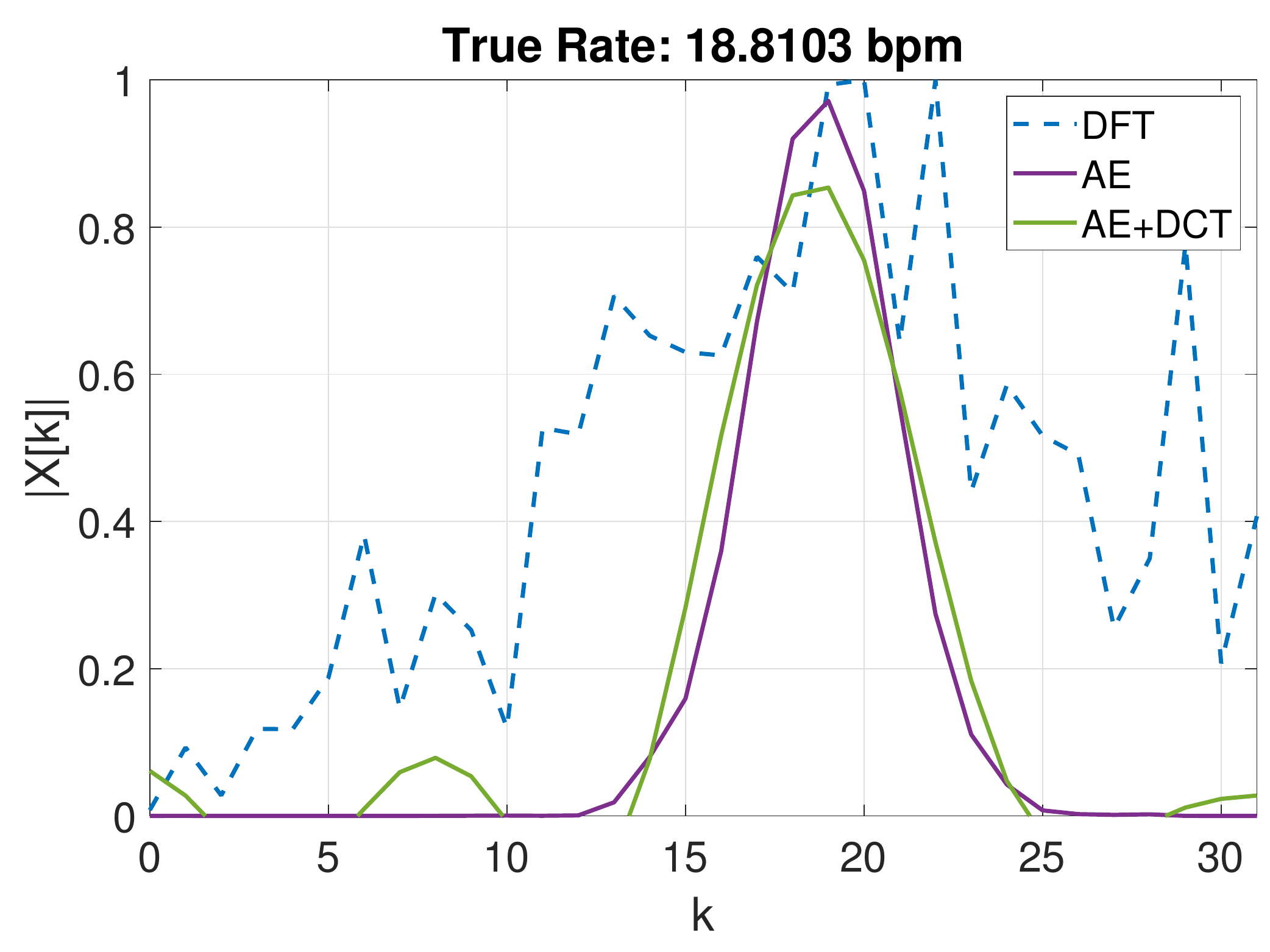}}\\
\caption{Outputs of autoencoder networks. In (a) and (b), both autoencoders with and without the DCT layer (purple and green) generate outputs with the correct peak value (the DFT spectrum is plotted in a dashed line). In (a) the plain autoencoder output is not as good as the proposed autoencoder with the DCT layer. In (b), DFT fails to generate a peak at the correct respiration rate.}
\label{fig: output}
\end{figure}

Typical autoencoder outputs are shown in Fig.~\ref{fig: output}. Regular autoencoder outputs are neither smooth nor as decisive as the outputs of the proposed autoencoder with the DCT layer generates smoother outputs as shown in Fig.~\ref{fig: output}.
For example, in Fig.~\ref{fig: output a}, the value of the index $k=12$ bpm is very close to the true value $k=13$ bpm. This is probably due to the peak at $k=10$ in the original ECG spectrum. To solve such cases we introduce an additional DCT layer to the network to generate smoother outputs clearly indicating the location of the respiration rate.
In the following subsection, we discuss the proposed DCT layer with trainable parameters.

\subsection{Autoencoder Network with a DCT Layer}\label{sec: Autoencoder Network with a DCT Layer}
In our previous work, we successfully used DCT layers in the convolutional neural networks for drift compensation in chemical sensors as described in~\cite{badawi2021discrete}. In this section, we design a trainable DCT network as shown in Fig.~\ref{fig: dct} for the autoencoder.  

In detail, the DCT layer is designed as follows:
First, a type-II DCT~\cite{ahmed1974discrete} is applied to the input:
\begin{equation}
    X_k = \sqrt{\frac{2}{N}}\sum_{n=0}^{N-1}x_n\text{cos}\left[\frac{\pi}{2N}(2n+1)k\right],\label{eq: DCT}
\end{equation}
for $0\leq k \leq M-1, M=8, N=32$. In the DCT domain, we keep the first $M=8$ DCT coefficients which is the dimension of the latent space. High-indexed DCT coefficients correspond to high-frequency components in the input and they correspond to noisy variations in the input, so they should be removed. This is a similar data processing approach as in signal and image coding, PCA, and autoencoders~\cite{pan2021robust, pan2022multiplication, pan2022detecting}. DCT approximates PCA matrices when there is a high correlation among the input samples \cite{ahmed1974discrete}. 

Then, each DCT coefficient is scaled by a trainable weight parameter $V_k$:
\begin{equation}
    \tilde{X}_k = V_k\cdot X_k,
\end{equation}
for $0\leq k \leq M-1, M=8$. The weight of each DCT coefficient is learned during the training of the network using the backpropagation algorithm.

Next, a trainable soft-thresholding layer~\cite{pan2021fast,pan2022block,pan2022deep} is applied to remove the small entries in the DCT domain similar to image coding and transform domain denoising. Soft-thresholding denoises the ECG data as well. We determine the thresholds using the backpropagation algorithm. The trainable soft-thresholding layer is defined as follows:
\begin{equation}
    \hat{X}_k = \mathcal{S_T}(\tilde{X}_k) = \text{sign}(\tilde{X}_k)(|\tilde{X}_k|-T_k)_+,
\end{equation}
where $T_k$ is a trainable threshold parameter, $0\leq k \leq M-1, M=8$, and $()_+$ is the ReLU function. Nonlinear soft-thresholding functions are shown in Fig.~\ref{fig: dct}.

Finally, we compute $N=32$-point inverse-DCT (IDCT) to get the pseudo-spectrum:
\begin{equation}
    \hat{x}_n = \sqrt{\frac{2}{N}}\left(\frac{1}{2}\hat{X}_0+\sum_{k=1}^{M-1}\hat{X}_k\text{cos}\left[\frac{\pi}{2N}(2n+1)k\right]\right),\label{eq: IDCT}
\end{equation}
for $0\leq n \leq N-1, M=8, N=32.$ 

\begin{figure}[htbp]
\centering
\includegraphics[width=1\linewidth]{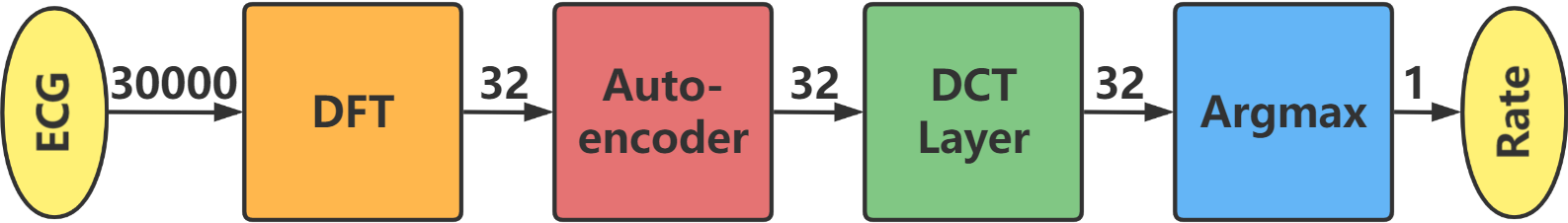}
\caption{The block diagram of the ECG-based respiration rate estimator. The numbers above the arrows denote the length of the tensors (vectors).}
\label{fig: model}
\end{figure}

In summary, the structure of the proposed ECG-based respiration rate estimator is shown in \ref{fig: model}. 

\section{Experimental Results}
We have collected 267 ECG measurements using the wireless ECG sensor device. The duration of each measurement is 60 seconds. The sampling rate is 500 Hz. We split the dataset into 167 training samples and 100 test samples. To train the neural networks, we use mean squared error (MSE) as the loss function and use AdamW optimizer~\cite{loshchilov2018decoupled} with the learning rate of 0.001 for 10000 epochs. During the training, the best models are saved based on their test MSE values. Evaluation for the MSE and mean absolute error (MAE) on the MSE-best models is reported in Table~\ref{tab: experimental results}. To reduce randomness in the results, we differently split the dataset three times and repeat the experiment. The last two rows indicate the average of the three experiments. Our experiments are implemented using PyTorch in Python 3. 
\begin{table}[htbp]
\centering
\begin{tabular}{ccccc}
\hline\noalign{\smallskip}
\bf{Split}&\bf{Metric} & \bf{DFT}& \bf{AE} & \bf{AE+DCT}\\
        \noalign{\smallskip}\hline\noalign{\smallskip}
\multirow{2}{*}{1}       & MSE    &0.8142 & 0.2907 & \bf{0.1205}\\
        & MAE    &0.6829 & 0.3586 & \bf{0.2802}\\
        \noalign{\smallskip}\hline\noalign{\smallskip}
\multirow{2}{*}{2}       & MSE    & 0.7445 & 0.2598 & \bf{0.1812}\\
        & MAE    &0.6530 & 0.3300 & \bf{0.3274}\\
        \noalign{\smallskip}\hline\noalign{\smallskip}
\multirow{2}{*}{3}       & MSE    & 0.7276 & 0.2359 & \bf{0.1555}\\
        & MAE    & 0.6387 & 0.3694 & \bf{0.3000}\\  
        \noalign{\smallskip}\hline\noalign{\smallskip}
\multirow{2}{*}{Average} & MSE    & 0.7621 & 0.2621 & \bf{0.1524}\\
        & MAE    & 0.6582 & 0.3527 & \bf{0.3025}\\  
        \noalign{\smallskip}\hline\noalign{\smallskip}
\end{tabular}
\caption{Experimental results. DFT is the traditional Fourier analysis method. ``AE" stands for the autoencoder-only RR estimation method. ``AE+DCT" is the proposed method of an autoencoder with a DCT layer.}
\label{tab: experimental results}
\end{table}

The accuracy results of the DFT-based peak detection method are improved by the classical autoencoder and the best results are achieved by the proposed autoencoder with a DCT layer. As shown in Table~\ref{tab: experimental results}, the autoencoder reduces the average MSE from 0.7621 to 0.2621 (65.58\%) and the average MAE from 0.6582 to 0.3527 (46.41\%), respectively. When we cascade the output of the autoencoder to the DCT layer, the average MSE is further reduced to 0.1524 (80.00\% less compared to the traditional analysis, 41.85\% less compared to the autoencoder), and the average MAE is further reduced to 0.3025 (54.04\% less compared to the traditional analysis, 14.23\% less compared to the autoencoder).

\section{Conclusion}
In this paper, we proposed a novel real-time Respiration Rate (RR) estimation method from ECG data obtained using an autoencoder with a DCT layer. Because of the measurement noise and uneven breathing, we may observe multiple peaks in the traditional DFT spectrum. As a result, a DFT-only based RR estimation may lead to inaccurate results. However, the autoencoder with a DCT layer can learn from training data and generate a  pseudo-frequency spectrum with a single peak. We experimentally observed that the accuracy of the RR detector can be improved significantly using an autoencoder.

We introduced a novel autoencoder structure with a DCT layer with trainable parameters in the DCT domain. The DCT layer further improves the results of the regular autoencoder. This is because the DCT decorrelates the input vector in the DCT domain (or latent space) similar to the Principal Component Analysis (PCA).
\small
\bibliographystyle{unsrt}
\bibliography{ref}

\end{document}